# Measuring bulk and surface acoustic modes in diamond by angle-resolved Brillouin spectroscopy

Ya-Ru Xie[1,2] Shu-Liang Ren[1,2] Yuan-Fei Gao[1,3] Xue-Lu Liu[1,2] Ping-Heng Tan[1] Jun Zhang[1,2,3,4,*]

[1] State Key Laboratory of Superlattices and Microstructures, Institute of Semiconductors, Chinese Academy of Sciences, Beijing, China

[2] Center of Materials Science and Optoelectronics Engineering, University of Chinese Academy of Sciences, Beijing, China

[3] Beijing Academy of Quantum Information Science, Beijing, China

[4] CAS Center of Excellence in Topological Quantum Computation, University of Chinese Academy of Sciences, Beijing, China

**Correspondence**

Jun Zhang, State Key Laboratory of Superlattices and Microstructures, Institute of Semiconductors, Chinese Academy of Sciences, Beijing, 100083, China.

Email: zhangjwill@semi.ac.cn

**Abstract**

The acoustic modes of diamond not only are of profound significance for studying its thermal conductivity, mechanical properties and optical properties, but also play a determined role in the performance of high-frequency and high-power acoustic wave devices. Here we report the bulk acoustic waves (BAWs) and surface acoustic waves (SAWs) of single crystal diamond by using the angle-resolved Brillouin light scattering (BLS) spectroscopy. We identify two high-velocity surface skimming bulk waves, with sound velocities of $1.277 \times 10^6$ and $1.727 \times 10^6$ cm/s, respectively. Furthermore, we conduct the relationship among the refractive index, incident angle and the velocities of BAWs propagating along an arbitrary direction. Our results may provide a valuable reference for fundamental researches and devices engineering in the community of diamond-based acoustic study.

## 1 INTRODUCTION

The acoustic waves, consisting of bulk acoustic wave (BAW) and surface acoustic wave (SAW), are caused by density fluctuations in matters and provide information about the elasticity, electrostriction, and thermal capacity of materials[1-4]. Both the SAW-based and BAW-based devices have been widely applied in the communication and sensing devices, such as acoustic radio frequency (RF) filter, resonator and SAW microelectromechanical systems (MEMS)[5-8]. The use of micro-acoustic devices greatly shrunk the size of signal processor in the field of communications. With the rapid development of the mobile communications, more high-frequency and high-power acoustic wave devices will be needed[9].

Diamond shows excellent performance in the high-frequency and high-power acoustic wave devices, thanks to its superhigh elastic modulus and thermal conductivity, stable chemical property as well as low thermal expansion coefficient and dielectric constant[10]. The piezoelectric film/diamond multilayer structure has been extensively applied in high-frequency and high-power SAW devices[11-13]. In addition, the emergent quantum acoustodynamics (QAD) cavities based on diamond make a contribution for the development of the hybrid quantum devices[14,15]. Since the spins of single negatively charged silicon vacancy (SiV) center[16,17] and nitrogen vacancy (NV) center[18, 19] have a long coherence time, these defects have been demonstrated to achieve coherent interaction with SAWs. Therefore, further research on the acoustic properties of diamond is of great significance for understanding its basic physical properties and designing high-performance acoustic wave devices.

Inelastic light scattering[20] is a precise method to detect phonon in solid materials. To measure the low-frequency acoustic phonon spectra of diamond, Brillouin light scattering is used in this work. As a noncontact and powerful method, the high-resolved Brillouin spectroscopy has been widely used to detect the acoustic waves in the frequency below 300 GHz[4]. There are three mechanisms contributing to the BLS process: scattering from the BAWs through the elasto-optic mechanism, scattering from the SAWs via the "surface ripple" mechanism, and scattering from the SAWs with phase velocities consistent with BAWs[21]. In the past few decades, several different techniques have been used to investigate the BAWs of diamond, such as inelastic neutron scattering[22], inelastic synchrotron scattering[23], and ultrasonic-pulse technique[24]. Beyond these techniques, BLS spectroscopy also was widely used to study the acoustic waves in diamonds. For example, Grimsditch and Ramdas have measured the BAWs along the high symmetry direction of diamond and studied their elastic moduli and elasto-optic constants[25], but they didn't give any angle dependent information and dispersion of BAWs along non-high-symmetry direction of diamond. Motochi et al obtained two SAW-like modes of CVD diamond with BLS experiment[26], but they did not explain origins of these SAW-like modes. Therefore, it is necessary to further explore angle dependent properties of the BAWs and investigate the physical mechanism of SAW of diamond. Such a study should benefit the development of acoustic devices based on diamond.

Here we report the angle-resolved BLS spectra of the BAWs and SAWs of a diamond single-crystalline sample [(100) oriented] placed on a pristine silicon wafer [(111) oriented]. This work aims to explore the properties of diamond acoustic mode dependent on the angle of incident light, and explain the origins of two high-speed surface acoustic modes. With the increasing incident angle, the longitudinal acoustic (LA) peak shows a blueshift trend, while the transverse acoustic (TA) peak gradually splits into two branches. We attribute this phenomenon to the velocity diversity of the acoustic mode along different crystal directions. Furthermore, we also identify three modes like SAW: Rayleigh SAW (RSAW), surface skimming transverse wave (SSTW) and surface skimming longitudinal wave (SSLW) , whose velocities are $1.080 \times 10^6$ cm/s, $1.277 \times 10^6$ and $1.727 \times 10^6$ cm/s, respectively.

## 2 EXPERIMENTAL DETAILS

Our experimental setup is shown in Figure 1. The Brillouin spectra of diamond were measured by a confocal microscopic BLS system in the backscattering geometry. The system is composed of high contrast (about $10^{15}$) (3+3)-pass tandem Fabry-Pérot interferometers (FPI) and the confocal microscope (CM) with a 20× bright field objective lens (numerical aperture NA = 0.42), both from the JSR Scientific Instruments. The detector is the Hamamatsu H10682-110 with a quantum efficiency of 10.8%. When the laser passes to the polarizing beam splitter (PBS), the *s*-polarized light is almost totally reflected to reach the sample, while the *p*-polarized component is nearly completely transmitted. The incident light, emitted from a single longitudinal mode laser source with a wavelength of 532 nm, is focused on the top surface of diamond. It assures that the collected scattering information is from as near to the surface as possible due to the confocal setup. The laser power is 29 mW, and we did not observe any laser heating effect on the sample owing to the high thermal conductivity of diamond. In our experiments, there are three kinds of polarization configuration: circular polarization ($\sigma^+ \sigma^-$), parallel polarization (VV) and cross polarization (VH), corresponding to only the quarter wave plate (QWP) in the light path, both the QWP and polarizer in the light path as well as neither the QWP nor polarizer in the light path, respectively. Here, the first and second items of the polarization abbreviation describe the polarization of incident light and scattered light. Specifically, $\sigma^+$ ($\sigma^-$), V and H represent right (left) circular-polarization, *s*-polarization and *p*-polarization, respectively.

The diamond with a size of 3×3×0.25 mm³, purchased from the Element Six Technologies, is a type IIa single crystal synthesized by CVD method. It has two (100) oriented surfaces which are polished with a roughness <30 nm and its 12 edges are along the <100> orientation with a miscut within 3°. The diamond is placed on a silicon wafer fixed on the home-built angle-resolved holder and can be rotated around the *z*-axis, as shown in Figure 1. The incident angle $\theta_i$ is adjusted by rotating the diamond to achieve angle-resolved BLS measurements. The rotation accuracy is 1°. The parameters of diamond are: density $\rho$ = 3.515 g/cm³, refractive index *n* = 2.426 when $\lambda$ = 532 nm.

## 3 RESULTS AND DISCUSSIONS

The BLS is an inelastic scattering of light from acoustic phonon due to the density fluctuations of materials [27-29]. The Brillouin shift, the frequency *f* of scattered acoustic wave is written as:

$$f = qv, \qquad (1)$$

where, $v = \sqrt{X/\rho}$ is the acoustic velocity, *X* is an expression composed by the elastic modulis $C_{ij}$, $\rho$ is the density of sample; *q* presents the magnitude of wave vector ***q***.

In a BLS experiment, the wave vector of scattering light $k_s$ and scattering angle $\alpha$ are defined by scattering geometry. For a given wave vector of incident light $k_i$, the detected wave vector of BAW is $q = \pm 2n k_i \sin\frac{\alpha}{2}$; the wave vector of SAW is $q^{\parallel} = k_i \sin\theta_i + k_s \sin\theta_s$, where the $\theta_i$ is the incident angle and the $\theta_s$ is the angle between the direction of the collected scattering light and the normal vector of sample surface. In backscattering geometry, there are $\theta_s = \theta_i$ and $k_s = k_i$. Therefore, ***q*** and $q^{\parallel}$ are equal to $\pm 5.73 \times 10^5$ cm⁻¹ and $\pm 2.36 \sin\theta_i \times 10^5$ cm⁻¹ in our experiments, respectively. In other words, the wave vector of BAW in the sample is a fixed value and does not depend on the incident angle, while the wave vector of SAW is proportional to $\sin\theta_i$. In fact, since the numerical aperture of the objective lens and the incident aperture of the interferometer are not negligible, the $k_i$ and $k_s$ have a certain uncertainty $\Delta k$ instead of a single value, which makes $q^{\parallel}$ also have an small uncertainty $\Delta q$. This wave vector uncertainty has almost no effect on BAW due to ***q*** >> $\Delta q$, but it has a significant effect on the acoustic mode propagating near the surface. When $\theta_i = 0$ (i.e. vertical incidence), $q^{\parallel} \sim \Delta q$, so the measured SAW frequency has a certain range instead of zero, and it appears as a broadening of the Rayleigh line on the spectra; as the incident angle becomes larger, $q^{\parallel}$ increases while the value of $\Delta \boldsymbol{q}$ does not change. When $q^{\parallel}$ is much larger than $\Delta \boldsymbol{q}$, the broadening effect caused by $\Delta \boldsymbol{q}$ becomes ignorable, and the full-width at-half-maximum (FWHM) of SAW decreases when the $\theta_i$ approaches 90°. Thus, by choosing a smaller numerical aperture of objective lens and a smaller entrance aperture of FPI, this broadening effect can be effectively suppressed, but the detection efficiency will be reduced simultaneously. In actual experiments, we always make a compromise between the aperture and detective efficiency.

For an arbitrary propagation direction, there are two quasi-TA modes and one quasi-LA mode. The LA usually propagates faster than TA[1]. In the Table 1, we list the reported acoustic velocities of diamond propagating along three high symmetry axes and our measured results. In the BLS spectra, the intensity and polarization properties of scattering signals are determined by the scattering tensors of acoustic modes. Scattering tensors are dependent on the direction of wave vector $\boldsymbol{q}$ of scattered acoustic phonons. For the $\boldsymbol{q}$ paralleling to [100] orientation, the scattering tensors **T** of BAWs along three different vibration directions are:

$$\text{LA:  } \boldsymbol{u} = [100] \quad \mathbf{T} = \varepsilon_0^2 \begin{bmatrix} p_{11} & 0 & 0 \\ 0 & p_{12} & 0 \\ 0 & 0 & p_{12} \end{bmatrix} \quad (2)$$

$$\text{TA}_1\text{:  } \boldsymbol{u} = [001] \quad \mathbf{T} = \varepsilon_0^2 \begin{bmatrix} 0 & 0 & p_{44} \\ 0 & 0 & 0 \\ p_{44} & 0 & 0 \end{bmatrix} \quad (3)$$

$$\text{TA}_2\text{:  } \boldsymbol{u} = [010] \quad \mathbf{T} = \varepsilon_0^2 \begin{bmatrix} 0 & p_{44} & 0 \\ p_{44} & 0 & 0 \\ 0 & 0 & 0 \end{bmatrix} \quad (4)$$

Here $\boldsymbol{u}$ is the vibration direction of phonons traveling along $\boldsymbol{q}$, $\varepsilon_0$ is the dielectric constant of material and $p_{ij}$ is the elasto-optic constants. Under a given polarization configuration, the intensity $I$ of scattered peak is proportional to $[\hat{e}_s \cdot \mathbf{T} \cdot \hat{e}_i]^2$, where $\hat{e}_s$ and $\hat{e}_i$ are unit vectors along the polarization direction of the scattered and incident light, respectively. For the $\boldsymbol{q}$ paralleling to [100], the $[\hat{e}_s \cdot \mathbf{T} \cdot \hat{e}_i]^2$ of LA mode is equal to $p_{12}^2$ under VV polarization, while both two TA modes are zero intensity; under VH polarization, only the $[\hat{e}_s \cdot \mathbf{T} \cdot \hat{e}_i]^2$ of TA$_2$ mode is not zero. For $\sigma^+\sigma^-$ polarization, all modes are not zero.

Figure 2a to Figure 2c display the angle-resolved BLS spectra of diamond under three polarization configurations. As mentioned above, LA mode is invisible in VH polarization and TA mode is invisible VV polarization. When the laser is perpendicularly incident on the sample surface (i.e. $\theta_i = 0°$), the measured frequencies of LA and TA modes are at ~159 and ~116 GHz, respectively. Based on the measured frequency, we find that their velocities are consistent with LA and TA modes along Γ-X direction as shown in Table 1. With the $\theta_i$ increasing, the frequency of LA mode has a blueshift, and the TA mode gradually splits into two peaks: one has an unchanged frequency at 116 GHz and the other has a redshift, as shown in Figure 2d to 2g. The velocities of TA and LA mode are also listed in Table 1. The A, B and C modes are not observed at $\theta_i = 0°$, but they become visible with increasing $\theta_i$ and their frequencies are proportional to $\theta_i$ as shown in Figure 2h. In Figure 2i, we can see that the A and B modes are very close together and almost merge into one broad peak, and it has VH polarization. Since spurious signals affect the analysis of peaks in 30-40 GHz in VV and σ⁺σ⁻ configurations as shown in figure 2a and 2c, here we only analyze spectra of A and B modes under VH polarization.

Figure 3a displays the 1/8 zone of first BZ of the diamond[31]. Since the point group of the diamond belongs to $O_h$ with 48 symmetry element[32], the first BZ can be divided into 48 equivalent irreducible wedges (IW)[33]. The IW shown as the shaded area in Figure 3a, is defined by the primitive vectors paralleling to three high symmetry axes of BZ, Γ-X, Γ-K and Γ-L, respectively. Each IW contains all of BAWs in diamond. According to our experimental geometry, the wave vectors of incident light and scattering light vary in the x-o-y plane of the IW shown in Figure 3b. Therefore, the wave vector of scattered phonons is a linear combination of $q_{ΓX}$ and $q_{ΓK}$ along the high symmetry axes. The measured BAWs propagate along or against the direction of refracted light (Γ-A direction) with a velocity of $v_{ΓA}$. The obtained velocities of TA and LA modes along Γ-A direction are listed in the Table 1. As shown in Figure 3b, $v_{ΓA}$ is the vector sum of $v_{ΓX}$ and $v_{ΓK}$ in the x-o-y plane of IW. In a parallelogram AСΓD with a 45° angle, we can easily find that:

$$v_{ΓA} = v_{ΓX}\frac{q_{ΓX}}{q}\cos\theta_i' + v_{ΓK}\frac{q_{ΓK}}{q}\cos(45°-\theta_i'), \tag{5}$$

where $q_{ΓX}$ and $q_{ΓK}$ are two components of $q$ propagating along Γ-X and Γ-K direction, respectively. In fact, it is more concise to obtain the velocities by using the incident angle rather than the refractive angle:

$$v_{\Gamma A} = v_{\Gamma X} \frac{\sqrt{n^2 - \sin^2\theta_i}}{n^2}(\sqrt{n^2-\sin^2\theta_i}-\sin\theta_i) + v_{\Gamma K} \frac{\sin\theta_i}{n^2}(\sqrt{n^2-\sin^2\theta_i}+\sin\theta_i), \qquad (6)$$

where, $\theta_i$ is the incident angle between 0° and 90°, and the direction of diamond rotating around the z-axis cannot affect its sign. When $\theta_i = 90°$, the $\theta_i'$ has a maximum $\theta_i' = 24.3°$ based on Snell's law. Thus, we cannot measure the LA and TA modes traveling along the Γ-K direction. By establishing the connection between the high symmetry axes in BZ and the arbitrary direction in crystal, the equation 6 provides a convenient method to calculate the velocity of BAW propagating along an arbitrary direction.

In Figure 3c, we plotted the velocities of three BAWs as function of $\sin\theta_i$ and corresponding fitting curves by equation 6. Considering that the LA mode is single-peak, its error bars are from the measurement resolution, while the error bars of two TA modes are from fitting error of the double-peak fitting. The velocities of BAW modes along [110] orientation are obtained by fitting: the LA modes is $1.808\times10^6$ cm/s, the two TA modes are $1.268\times10^6$ cm/s and $1.170\times10^6$ cm/s, respectively. It is coinciding with the previous reports and we list these data in Table I. According to $f = \pm 5.73\times10^5 v$, we plotted the frequency curves depending on $\sin\theta_i$ in Figure 3d and got the same conclusion. That is, the velocity of LA (TA) mode propagating along an arbitrary direction can be calculated with the vector sum method in IW. In our experiments, the propagating direction of BAWs gradually change from Γ-X to Γ-K direction as the $\theta_i$ increases. As shown in Table 1, for LA mode, $v_{\Gamma X} < v_{\Gamma K}$; while for TA mode, $v_{\Gamma K}$ has two different values in two orthogonal vibration directions, one equals to $v_{\Gamma X}$ and the other is smaller. That is why the LA peak is blue-shifted, the TA peak is split and one of TA peaks is redshift during the rotation of diamond sample.

Compared with BAWs, SAWs are typical acoustic modes propagating along the surface. Only when the incident light wave vector have a projection on the surface, $q^{\parallel}$ is non-zero, and SAWs can be observed. Under our angle-resolved BLS configuration, the $q^{\parallel}$ is gradually growing as the angle $\theta_i$ increasing, and the dispersion relation of SAWs can be wrote as $f = q^{\parallel}v = \pm 2.36\times10^5 \sin\theta_i v$. We found that A, B and C modes have the characteristics of SAW: their frequencies approache to zero when $\theta_i = 0°$; besides, their frequencies are obviously proportional to $\sin\theta_i$ as shown in Figure 4b. Entropy fluctuation is a possible reason to cause the widening of Rayleigh line, so-called Rayleigh-center scattering[34], but it cannot explain the phenomenon that the split-up of Rayleigh line and the blue shift of these modes. As mentioned above, the SAW peaks are broadened due to the finite numerical aperture of objective lens and entrance aperture of interferometer under a small $\theta_i$. That is why the Rayleigh peak is wider when $\theta_i = 0°$ than $\theta_i > 40°$ in Figure 2b. We use above dispersion relationship of SAW to linearly fit the experimental data of these three modes, and obtain their velocities: $v_A = (1.277 \pm 0.011) \times 10^6$ cm/s, $v_B = (1.080 \pm 0.009) \times 10^6$ cm/s and $v_C = (1.727 \pm 0.010) \times 10^6$ cm/s.

The velocity of B mode meets the conditions of RSAW who possesses a phase velocity about 10% lower than that of the slower BAW (TA2 in this work) on the same substrate surface. RSAW is mainly caused by the couple of longitudinal (L-) and shear vertical (SV-) type BAW close to the surface, so most of the energies concentrates near the surface[35], as schematically shown in Figure 4a. Jiang et al reported that the calculated velocity of RSAW is 10753 m/s and measured value is 10326±470 m/s using stimulated Brillouin scattering on polycrystalline CVD diamond film[36]. Djemia et al [37] also reported that RSAW is 10800 ± 300 m/s measured by BLS. Our result of $v_B = (1.080 \pm 0.009) \times 10^6$ cm/s is consistent with their results, so we attribute B mode to the RSAW.

Notably, the velocities of A and C are close to that of TA and LA mode, respectively, rather than smaller than the slower TA mode. We calculated the refractive index $n_{LA} = 2.483$ and $n_{TA} = 2.419$ when $\theta_i = 60°$, respectively, according to $n = f_{BAW}/f_{SAW} \sin\theta$, $f_{LA}$=161.92 GHz and $f_{TA1}$=115.76 GHz. It is consistent with the two high-speed SAW-like modes reported by Motochi et al[26]. Here, we attribute A and C mode as shallow bulk acoustic waves, which are also called pseudo-surface acoustic waves[38], or surface skimming bulk wave (SSBW)[35] in different applications fields. SSBW originates from the components of BAWs propagating along the surface, but its energy gradually leaks into the bulk at a propagation angle (the angle between surface and the propagation direction) smaller than leaky wave[39]. SSBW has two vibration configuration, one is transverse wave (SSTW) and the other is longitudinal wave (SSLW). There are many SSBWs with different velocities in a plate-shaped material, and their velocities are related to the thickness of the plate and the thickness of the dielectric layer on the plate[38]. Here, the measured SSBW could be from the interface between air and diamond layers. Since the incident light is focused on the diamond surface, the velocity of SSBW should be between that of TA mode and LA mode. Exactly, the velocities of A and C modes are almost the same as the velocities of TA mode and LA mode of diamond, respectively. Moreover, the polarization property of A mode is similar with TA mode, as shown in Figure 2a and 2b. Thus, the A mode is attributed to SSTW, whose displacement direction of particle (DDP) is parallel to the surface of sample but perpendicular to the propagation direction of wave; the B mode is attributed to the SSLW[40] , where the DDP is parallel to both sample surface and the propagation direction of wave, as shown in Figure 4a. The propagation of SSBW on the substrates cannot be affected by the surface roughness, and it has been applied to fabricate the high frequency and temperature stable devices due to its high propagation velocity[41-43].

**CONCLUSION**

In this study, we investigate the acoustic velocity propagating along an arbitrary direction inside the crystal based on angle-resolved BLS spectroscopy. We give the acoustic velocities of BAWs as a function of incident angle and the refractive index. Since the angle-resolved BLS spectroscopy can measure the surface wave vector by changing angle of incident light, we have successfully detected three surface waves from the (100) face of a crystalline diamond and discussed their physical mechanisms. Furthermore, it makes possible to design higher frequency diamond-based surface acoustic devices because SSBW has a higher propagation velocity than RSAW.

**ACKNOWLEDGEMENTS**

The authors thank the support from National Basic Research Program of China (grant no. 2016YFA0301200, 2016YFA0300804), Beijing Natural Science Foundation (JQ18014), Strategic Priority Research Program of Chinese Academy of Sciences (grant no. XDB28000000) and NSFC (12074371, 51527901).

TABLE 1 The sound velocity of diamond

| Propagation direction of $q$ | Velocity of Acoustic mode ($\times 10^6$ cm/s) | Previous reports | This work | Vibration direction of mode |
|---|---|---|---|---|
| [100] (Γ-X) | $v_{LA}$ | 1.751[a] | 1.745±0.0041 | [100] |
|  | $v_{TA}$ | 1.282[a] | 1.268±0.0003 | [010]/[001] |
| [110] (Γ-K) | $v_{LA}$ | 1.833[b] | 1.808±0.0038 | [110] |
|  | $v_{TA}$ | 1.282[b] | 1.268±0.0003 | [001] |
|  |  | 1.166[b] | 1.170±0.0001 | [1$\bar{1}$0] |
| [111] (Γ-L) | $v_{LA}$ | 1.857[a] |  | [111] |
|  | $v_{TA}$ | 1.208[a] |  | [1$\bar{1}$0]/[11$\bar{2}$] |
| [100]→[110] (Γ-A) | $v_{LA}$ |  | 1.745-1.790 |  |
|  | $v_{TA}$ |  | 1.268-1.218 |  |

[a] is calculated by $v_{LA} = \sqrt{C_{11}/\rho}$ and $v_{TA} = \sqrt{C_{44}/\rho}$ where $\rho$ = 3.512 g/cm³ is the density of diamond in this study. The elastic modulis $C_{44}$ and $C_{11}$ are cited from reference [25].

[b] Reference [24].

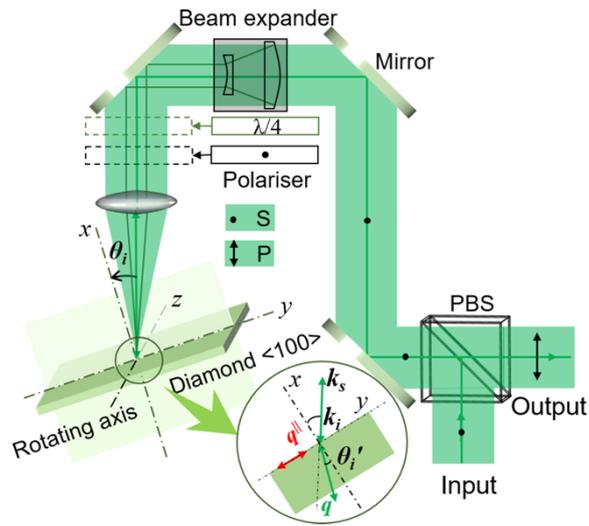

height=8cm, width=8cm

Figure 1 Experimental setup. Only the *s*-polarized component of incident light is completely reflected by the PBS and finally focuses on the diamond surface. The inset is a schematic diagram of light and phonon wave vectors involved on *x-o-y* plane, where phonons include SAW (wave vector is $q^{\parallel}$) and BAW (wave vector is $q$). The $\theta_i'$ is the refraction angle inside of the diamond.

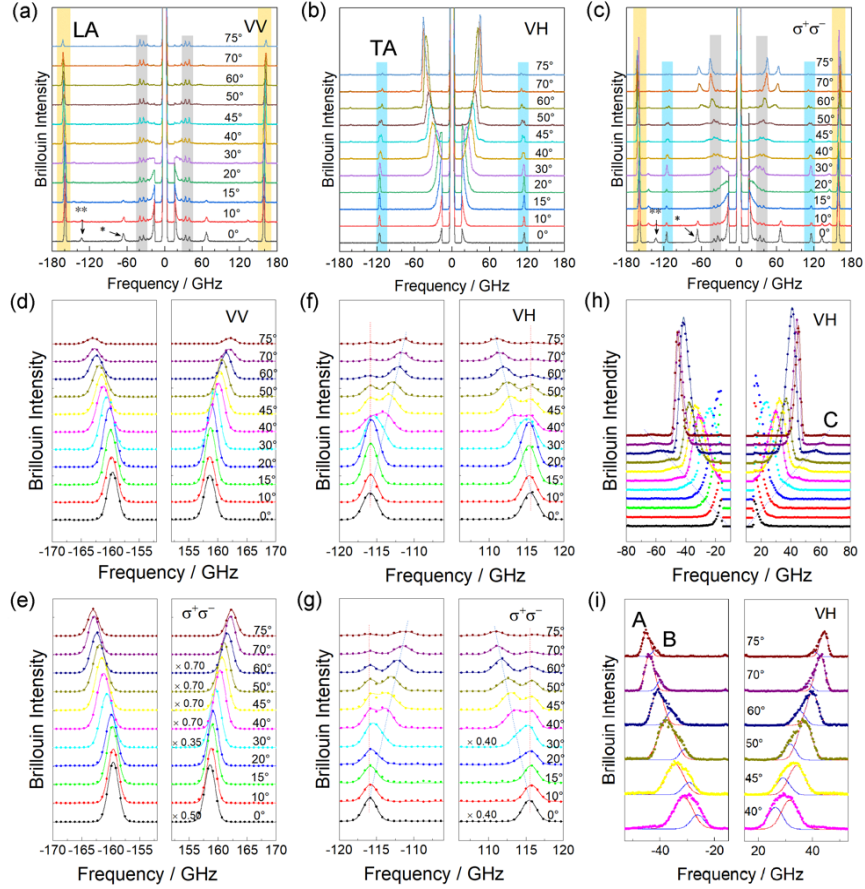

height=12cm, width=12cm

Figure 2 Angle-resolved BLS spectra of diamond. (a)-(c) The angle-resolved BLS spectra of diamond under three polarization configurations. The LA mode is marked with the yellow background and the TA mode is marked with the cyan background. The peaks marked with **, * and the two small sharp peaks with the gray background (frequency between 30-40 GHz) are spurious signals independent of samples. The frequency changes of (d)-(e) LA mode, (f)-(g) TA mode and (h) the RSAW (B) and two SSBWs (A and C) with the incident angle $\theta_i$. (i) Spectra of A and B modes measured at higher resolution. The solid circles represent the experimental data. The solid lines represent the fitting curves by Gauss function. The dash lines in (f) and (g) are the guides for the eyes.

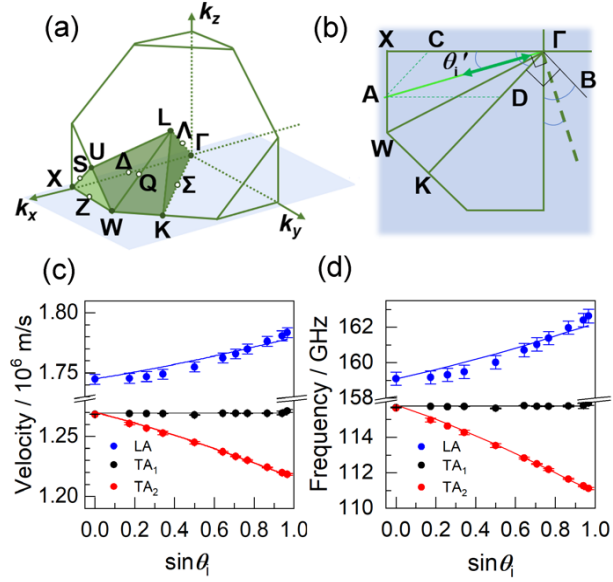

Figure 3 Schematic diagrams of (a) the 1/8 of the first BZ and (b) the *x-o-y* plane of the first BZ. The $\theta_i'$ corresponds to the refraction angle in real space. The green double-arrow indicates the wave vector of BAW. The dotted line is the auxiliary line for vector synthesis. (c) Velocity and (d) frequency of the BAWs of diamond as function of $\sin\theta_i$. The solid circles represent experimental data. The solid lines in are fitting curve based on equation (6) (c) and $f = qv$ (d).

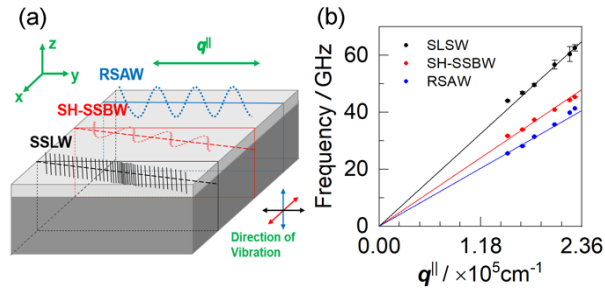

Figure 4 (a) Schematic diagrams and (b) the $f(q^{\|})$ relation of RSAW and two SSBWs. In (a), the area close to surface is filled with light gray, and the one far from the surface is filled with dark gray. The propagating direction of each mode is parallel to the plane shown by the block in corresponding colors. $q^{\|}$ is the projection of incident light wave vector on the surface. In (b), the solid circles represent experimental data and the solid lines are fitting curves by $f = \pm 2.36 \times 10^5 \sin\theta_i v$.